\documentclass[aps,prl,reprint,superscriptaddress,amsmath]{revtex4-1}
\usepackage{bbm}
\usepackage{verbatim}
\usepackage{graphicx}
\usepackage{times}
\usepackage{amssymb}
\usepackage{epsfig}
\usepackage{graphicx}
\usepackage{bm}
\usepackage{times}
\usepackage{txfonts}
\usepackage{dsfont}
\usepackage{algorithmic}
\usepackage{color}
\usepackage{textcomp}
\usepackage[unicode=true,pdfusetitle,
 bookmarks=true,bookmarksnumbered=false,bookmarksopen=false,
 breaklinks=true,pdfborder={0 0 0},backref=false,colorlinks=true]
 {hyperref}
\hypersetup{
    colorlinks,%
    citecolor=blue,%
    filecolor=blue,%
    linkcolor=blue,%
    urlcolor=blue
}

\begin{document}

\title{Simulating Quantum Fields with Cavity QED}

\author{Sean Barrett}
\affiliation{QOLS, Blackett Laboratory, Imperial College London, London SW7 2BW, United Kingdom}
\author{Klemens Hammerer}
\affiliation{Institute for Theoretical Physics, Leibniz University, 30167 Hannover, Germany}
\affiliation{Institute for Gravitational Physics, Leibniz University, 30167 Hannover, Germany}
\author{Sarah Harrison}
\affiliation{Department of Mathematics, Royal Holloway, University of London, Egham, TW20 OEX, United Kingdom}
\affiliation{Institute for Theoretical Physics, Leibniz University, 30167 Hannover, Germany}
\author{Tracy E. Northup}
\affiliation{Institut f\"ur Experimentalphysik, Universit\"at Innsbruck, 6020 Innsbruck, Austria}
\author{Tobias J. Osborne}
\affiliation{Institute for Theoretical Physics, Leibniz University, 30167 Hannover, Germany}

\date{\today}

\begin{abstract}
  As the realization of a fully operational quantum computer remains distant, \emph{quantum simulation}, whereby one quantum system is engineered to simulate another, becomes a key goal of great practical importance. Here we report on a variational method exploiting the natural physics of cavity QED architectures to simulate strongly interacting quantum fields. Our scheme is broadly applicable to any architecture involving tunable and strongly nonlinear interactions with light; as an example, we demonstrate that existing cavity devices could simulate models of strongly interacting bosons. The scheme can be extended to simulate systems of entangled multicomponent fields, beyond the reach of existing classical simulation methods.
\end{abstract}

\maketitle


Modelling interacting many-particle systems classically is a challenging yet tractable problem. However, in the quantum regime, it becomes rapidly intractable, owing to the dramatic increase in the number of variables required to describe the system. Feynman~\cite{Feynman1982} realized that an alternate approach would be to exploit quantum mechanics to carry out simulations beyond the reach of classical computers. This idea was the basis of Lloyd's simulation algorithm~\cite{Lloyd1996}, a procedure for a \emph{digital} quantum computer to simulate the dynamics of a strongly interacting quantum system. In contrast, there is also an \emph{analogue} approach to quantum simulation, where the simulator's Hamiltonian is tailored to match that of the simulated system~\cite{Buluta2009}. The complementary aspects of the analogue and digital methods, reviewed in \cite{Buluta2009, Aspuru-Guzik2012, Bloch2012, Johanning2009, Lewenstein2007}, have led to a host of recent experiments~\cite{Friedenauer2008, Gerritsma2010, Haller2010, Kim2010, Lanyon2010, Islam2011, Simon2011, Barreiro2011,Lanyon2011}.

To date, most experimental implementations of quantum simulation algorithms have been focussed on the task of simulating \emph{quantum lattice systems}, with comparatively less attention paid to systems with continuous degrees of freedom. The archetypal example of a quantum system with a continuous degree of freedom is the \emph{quantum field}. Currently, quantum simulations of quantum field theories have relied on discretization of the dynamical degrees of freedom. One body of recent theoretical work is focussed on the analogue simulation of discretized quantum fields, using cold atoms in optical lattices~\cite{Lepori2010,Bermudez2010,Cirac2010,Semiao2011,Kapit2011} and coupled cavity arrays~\cite{Hartmann2006,Greentree2006,Angelakis2007}. Complementing this are proposals for digital quantum simulation on a universal quantum computer of the zero-temperature~\cite{Byrnes2006} and thermal~\cite{Temme2011} dynamics of non-abelian gauge theories and, more recently, a digital quantum simulation~\cite{Jordan2012,Jordan2011} of scattering processes of a discretized $\lambda\phi^4$ quantum field.

In this paper we report on an \emph{analogue} algorithm to simulate the ground-state physics of a one-dimensional strongly interacting quantum field using the \emph{continuous}  output of a cavity-QED apparatus \cite{Raimond2001, Miller2005, Walther2006, Haroche2006, Girvin2009}. Our method involves no discretization of the dynamical degrees of freedom; the simulation register is the continuous electromagnetic output mode of the cavity. The variational wave function generated in this way therefore belongs to an extremely expressive class, namely the class of continuous matrix product states, as we will show. We argue that our approach is already realizable with state-of-the-art cavity-QED technology.

\begin{figure*}
\includegraphics[width=1.5\columnwidth]{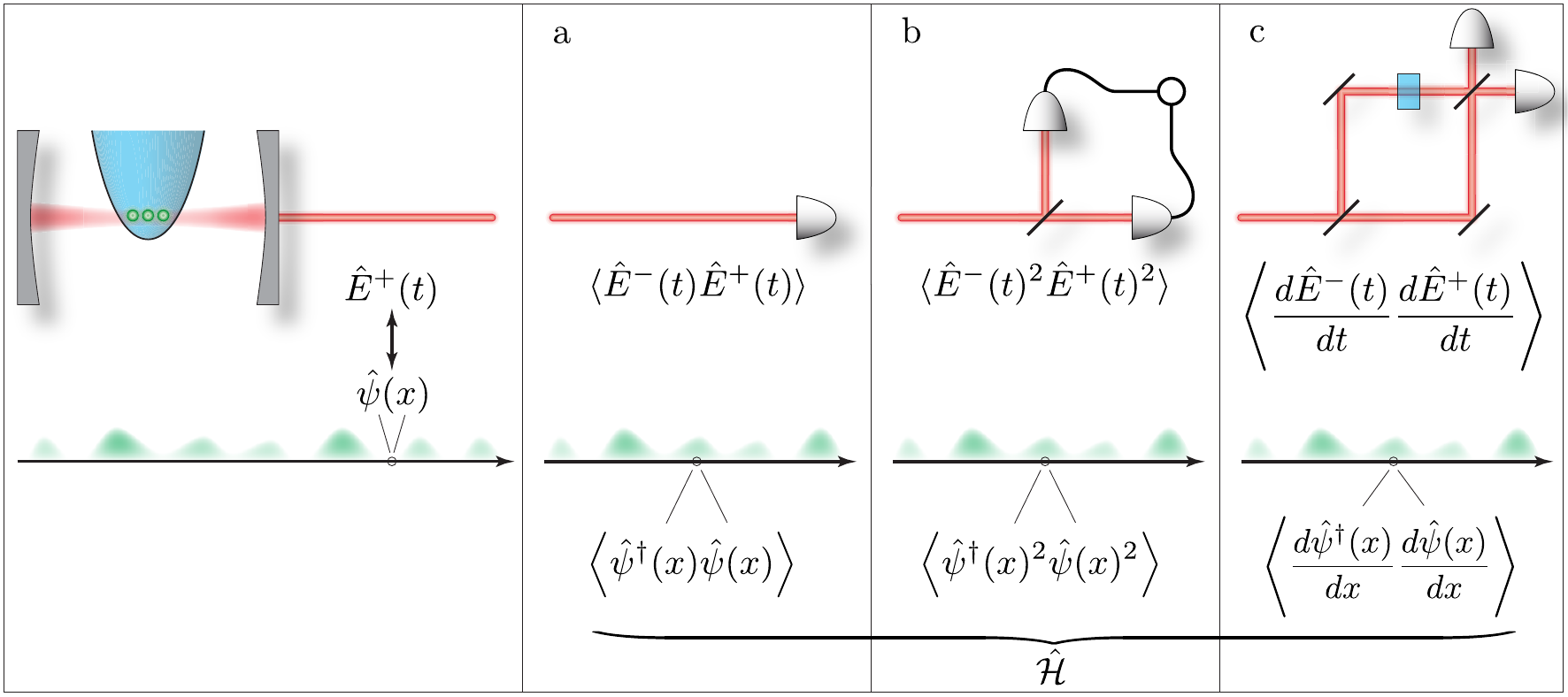}
\caption{The output field $\hat E^{+}(t)$ of a cavity-QED system is identified with a bosonic quantum field $\hat{\psi}(x)$.  Since optical detection schemes correspond to expectation values of quantum-field operators, $\langle \hat{\mathcal{H}} \rangle$ can be estimated via independent measurements of the cavity field.  For example, the operators $\hat{N}$, $\hat{W}$, and  $\hat{T}$ of Eq.~\eqref{eq:Ham} are determined, respectively, from measurements of (a) the output-field intensity, (b) Hanbury Brown and Twiss correlations, and (c) an interferometer with variable path length.}\label{fig:dictionary}
\end{figure*}


We concentrate on simulating quantum fields modelling collections of strongly interacting bosons in one spatial dimension.
These systems are compactly described in second quantization using the quantum field annihilation and creation operators $\hat{\psi}(x)$ and $\hat{\psi}^\dag(x)$, which obey the canonical commutation relations $[\hat{\psi}(x), \hat{\psi}^\dag(y)] = \delta(x-y)$.
The task is to determine the ground state of a given field-theoretic Hamiltonian $\hat{\mathcal{H}}(\hat\psi,\hat\psi^\dagger)$. The prototypical form of such a Hamiltonian is
\begin{equation}\label{eq:Ham}
	\hat{\mathcal{H}} = \int (\hat{T} + \hat{W} + \hat{N})\, dx
\end{equation}
where $\hat{T} = \frac{d\hat{\psi}^\dag(x)}{dx}\frac{d\hat{\psi}(x)}{dx}$, $\hat{W} = \int  w(x-y) \hat{\psi}^\dag(x)\hat{\psi}^\dag(y)\hat{\psi}(y)\hat{\psi}(x)\, dy$, and $\hat{N} = -\mu\hat{\psi}^\dag(x)\hat{\psi}(x)$ describe the kinetic energy, two-particle interactions with potential $w(x-y)$, and the chemical potential, respectively. Our approach provides a quantum variational algorithm for finding the ground states of an arbitrary Hamiltonian that is translation-invariant and consists of finite sums of polynomials of creation/annihilation operators and their derivatives.

The apparatus proposed to simulate the ground-state physics of $\hat{\mathcal{H}}$ is a single-mode cavity coupled to the quantum degrees of freedom of some intracavity medium (Fig.~\ref{fig:dictionary}); our proposal is not tied to the specific nature of the medium, so long as one or more tunable nonlinear interactions are present that are sufficiently strong at the single-photon level. Below we consider the example of a single trapped atom coupled to the cavity via electronic transitions. The system is described by a Hamiltonian $\hat H_{\mathrm{sys}}(\lambda)$ that depends on a set of controllable parameters $\lambda$, for example, externally applied fields. When the cavity is driven, either directly through one of its mirrors or indirectly through the medium, the intracavity field relaxes to a stationary state, and the cavity emits a steady-state beam of photons in a well-defined mode.

The crucial idea underlying our proposal is to regard the steady-state cavity output as a continuous register recording a \emph{variational} quantum state $|\Psi(\lambda)\rangle$ of a one-dimensional quantum field with control parameters $\lambda$ \emph{as} the variational parameters.
This representation is chosen so that the spatial location $x$ of the simulated translation-invariant field is identified with the value of the time-stationary cavity output mode exiting the cavity at time $t = x/s$. The arbitrary scaling parameter $s$ is included in the set of variational parameters $\lambda$.
We complete this identification by equating the annihilation operator $\hat{\psi}(x)$ of the simulated quantum field with the field operator $\hat{E}^{+}(t)$ for the positive-frequency electric field of the cavity output mode \footnote{We define $\hat E^{+}(t)$ in units such that \textlangle $\hat E^{-}(t)\hat E^{+}(t)$ \textrangle corresponds to the mean number of photons exiting the cavity per unit time.}, via $\hat{\psi}(x) =\hat E^{+}(t)/\sqrt{s}$.


Recall that the variational method proceeds by minimizing the average energy density of the variational state $f(\lambda) = \langle \Psi(\lambda)|\hat{T}+\hat{W}+\hat{N}|\Psi(\lambda)\rangle$ over the variational parameters $\lambda$. A key point in our scheme is that --- with the identification of the field operators $\hat E^{+}(t)$ and $\hat{\psi}(x)$ in hand --- the value of $f(\lambda)$ can be determined from standard optical measurements on the cavity output field, namely the measurement of \emph{Glauber correlation functions}~\cite{Glauber1963, Mandel1995}, see Fig.~\ref{fig:dictionary}. This result is easily seen for the Hamiltonian of Eq.~\eqref{eq:Ham}. Thanks to the linearity of the expectation value, we can separately measure $\langle \hat{T} \rangle$, $\langle \hat{W}\rangle$, and $\langle \hat{N} \rangle$.
The expectation value of the chemical
potential term corresponds to a function of the \emph{intensity}
of the output beam via $\langle \hat{N} \rangle  = -\frac{\mu}{s}\langle\hat E^{-}(t)\hat E^{+}(t)\rangle$. The kinetic energy term $\langle \hat{T} \rangle$ corresponds to the limit
\begin{align*}
\langle \hat{T} \rangle = \lim_{\epsilon_1,\epsilon_2\rightarrow 0} \frac{1}{s^3\epsilon_1\epsilon_2}&\left(g^{(1)}(t+\epsilon_1,t+\epsilon_2)-g^{(1)}(t+\epsilon_1,t)\right.\\
&\left.-g^{(1)}(t,t+\epsilon_2)+g^{(1)}(t,t)\right),
\end{align*}
where $g^{(1)}(t_1,t_2)=\langle\hat E^{-}(t_1)\hat E^{+}(t_2)\rangle$; this quantity can be estimated by choosing a finite but small value for $\epsilon_1$ and $\epsilon_2$.
Note that this procedure does not amount to a simple space discretization because the output is a continuous quantum register.
The final term $\langle \hat{W}\rangle$ depends on two-point spatial correlation functions $\langle \hat{\psi}^\dag(x)\hat{\psi}^\dag(y)\hat{\psi}(y)\hat{\psi}(x)\rangle$, which translate to measurements of
$g^{(2)}(t_1,t_2)=\langle\hat E^{-}(t_1)\hat E^{-}(t_2)\hat E^{+}(t_2)\hat E^{+}(t_1)\rangle$. The detection schemes to estimate all terms in the showcase Hamiltonian \eqref{eq:Ham} are presented in Fig.~\ref{fig:dictionary}.
From a wider perspective, \emph{any} Glauber correlation function $g^{(n,m)}=\langle\hat E^{-}(t_1)\cdots\hat E^{-}(t_n)\hat E^{+}(t'_m)\cdots\hat E^{+}(t'_1)\rangle$, i.e., any $n+m$-point field correlation function composed of $n$ creation operators $E^{-}$ and $m$ annihilation operators $E^{+}$~\cite{Glauber1963, Mandel1995}, can be measured with similar, albeit more complex setups, such as in~\cite{Gerber2009, Koch2011}.
Thus, upon identifying $E^{-}(t)$ and $E^+(t)$ with $\psi^\dagger(x)$ and $\psi(x)$, respectively, 
our scheme admits the measurement of any equivalent energy density $\propto \langle\hat\psi^\dagger(x_1)\cdots\hat\psi^\dagger(x_n)\hat\psi(x'_m)\cdots\hat\psi(x'_1)\rangle$, and therefore ultimately the simulation of arbitrary Hamiltonians $\hat{\mathcal{H}}(\hat\psi,\hat\psi^\dagger)$.


Once $f({\lambda})=\langle \Psi(\lambda)|\hat{\mathcal{H}}|\Psi(\lambda)\rangle$ has been experimentally estimated for a given $\lambda$,
the next step is to apply the variational method to minimize $f(\lambda)$.  Minimization is carried out by adaptively tuning the parameters $\lambda$ in the system Hamiltonian $\hat{H}_\mathrm{sys}(\lambda)$ and iteratively reducing $f(\lambda)$, for example, using a standard numerical gradient-descent method. Once the optimum choice of ${\lambda}$ is found, the resulting cavity output field is a variational approximation to the ground state of $\hat{\mathcal{H}}$, and relevant observables of the field theory can be directly measured using the 
detection schemes of Fig.~\ref{fig:dictionary}. We emphasize that our method applies also to cases where a numerical estimation of $f(\lambda)$ cannot be performed efficiently due to the size and complexity of the system to be simulated, and we suggest that this is exactly the strength of our approach. Moreover, the optimization may be performed experimentally without theoretically calculating the cavity-QED system dynamics; indeed, it is not necessary to accurately characterize $H_{\mathrm{sys}}$ or its relation to the adjustable parameters $\lambda$.

Why should the stationary output of a cavity-QED apparatus be an expressive class capable of capturing the ground-state physics of strongly interacting fields? It is possible to show that such states are of \emph{continuous matrix product state} (cMPS) type, a variational class of quantum field states recently introduced for the classical simulation of both nonrelativistic and relativistic quantum fields~\cite{Verstraete2010, Osborne2010, Haegeman2010a, Haegeman2010b}. These states are a generalisation of \emph{matrix product states} (MPS)~\cite{Fannes1992,Verstraete2008,Cirac2009,Schollwoeck2011}, which have enjoyed unparalleled success in the study of strongly correlated phenomena in one dimension in conjunction with the \emph{density matrix renormalization group} (DMRG)~\cite{White1992, Schollwoeck2005}. It turns out that \emph{all} quantum field states admit a cMPS description, providing a compelling argument for their utility as a variational class~\cite{Completeness}. Crucially, the cMPS formalism turns out to be \emph{identical} to the input-output formalism of cavity QED~\cite{Collett1984}. This identification was anticipated in~\cite{Verstraete2010,Osborne2010,Schoen2005,Schoen2007}, and we elucidate it further in the supplemental material. It implies that \textit{quantum field states emerging from a cavity are of cMPS type} and thus fulfill the necessary conditions for being a suitable and expressive class of variational quantum states.


\begin{figure}
\includegraphics[width=\columnwidth]{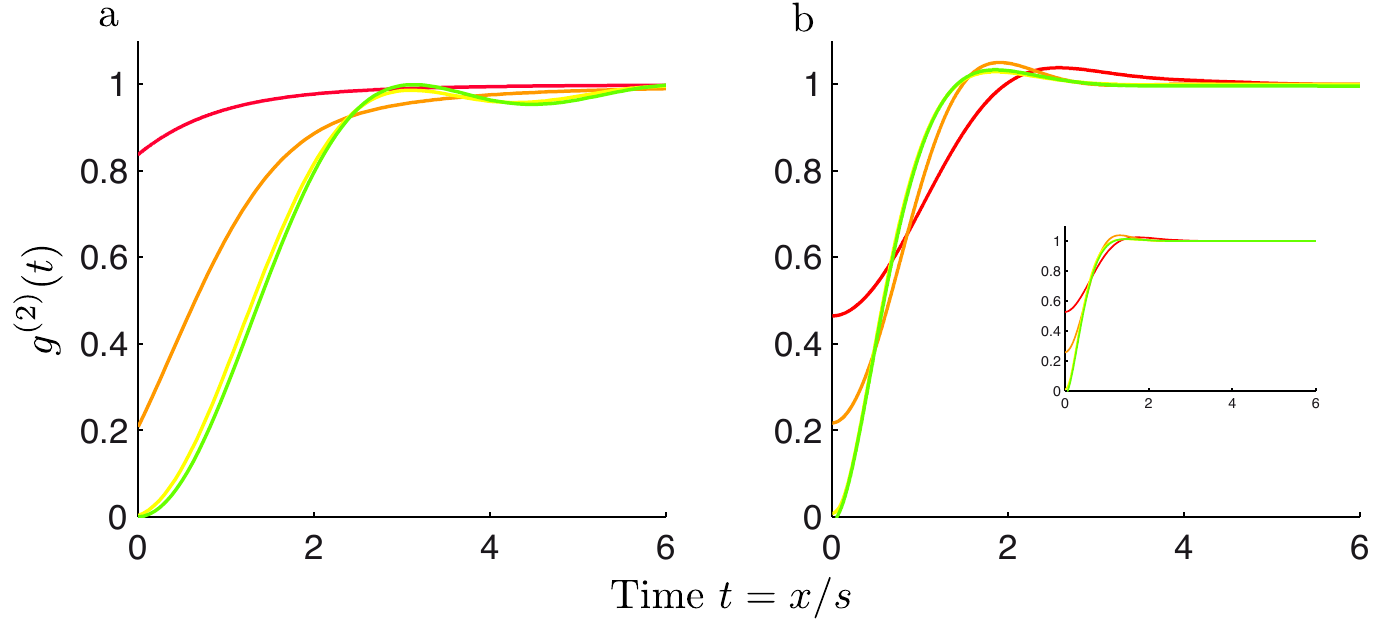}
\caption{Two-particle correlations in the Lieb-Liniger model are reproduced in simulations of an ion-trap cavity experiment. This \emph{critical} model exhibits a transition between the \emph{superfluid} regime for $v \approx 0$ and the \emph{Tonks-Girardeau} regime for $v\gg 0$, which is seen in the value of the correlation function at $t = 0$. (a) The Lieb-Liniger ground state is simulated for interaction strengths $v = \{ 0.07\, \text{(red)}, 3.95\, \text{(orange)}, 60.20\, \text{(yellow)}, 625.95\, \text{(green)}\}$, and correlation functions $\langle \hat{\psi}^{\dagger}(0)\hat{\psi}^{\dagger}(x)\hat{\psi}(x)\hat{\psi}(0) \rangle$ calculated as in \cite{Verstraete2010,Osborne2010} using 338 variational parameters.  (b) Two-photon correlation functions $g^{(2)}{(t)}$ for an experiment with the parameters of \cite{Stute2012}.  Although there are visible differences, with just three variational parameters $\{ g, \Omega, s \}$ the transition in the correlation functions is approximately reproduced. It is worth emphasizing how unusual it is for a variational calculation with only a few parameters to reproduce anything more than the coarsest features of a correlation function, e.g., if mean-field theory is used one does not obtain nontrivial correlators. Strikingly, the transition in (a) is captured even in the presence of realistic decay channels (inset). Note that this transition is analogous to that observed in \cite{Dubin2010}.}\label{fig:pilot_study}
\end{figure}

Even though cavity-QED output states are of cMPS type, can a realistic system in the presence of decoherence reproduce the relevant physics of a strongly interacting quantum field? As a test case, we demonstrate that the paradigmatic cavity-QED system, comprising a single trapped atom coupled to a high-finesse cavity mode, is capable of simulating the ground-state physics of an equally paradigmatic field, namely, the Lieb-Liniger model~\cite{Lieb1963}. This model describes hard-core bosons with a delta-function interaction and is given by Eq. \eqref{eq:Ham} with $w(x-y) = v\delta(x-y)$, where $v$ describes the interaction strength. Our simulator consists of a two-level atom interacting with one cavity mode, described (in a suitable rotating frame) by the on-resonance Jaynes-Cummings Hamiltonian
\begin{align}
\hat H_{sys} = g(\hat\sigma^{+}\hat a + \hat\sigma^{-}\hat a^{\dagger}) + \Omega(\hat\sigma^{+} + \hat\sigma^{-}),
\label{HJC}
\end{align}
where $\hat\sigma^{+}$ is the atomic raising operator and $\hat a$ is the cavity photon annihilation operator, $g$ the atom--cavity coupling,  and $\Omega$ the laser drive. The cavity-QED Hamiltonian $\hat H_{sys}$ can be realized in various experimental architectures~\cite{Raimond2001, Miller2005, Walther2006, Haroche2006, Girvin2009}. Here we choose the example of a trapped calcium ion in an optical cavity, with which tunable photon statistics have previously been demonstrated~\cite{Dubin2010}, and we show in the supplemental information (see below) how $g$ and $\Omega$ can function as variational parameters.

In an experiment, to measure the variational energy density $f(\lambda)$, the output beam would be allowed to relax to steady state, the intensity $I$, $g^{(1)}$, and $g^{(2)}$ functions estimated as depicted in Fig.~\ref{fig:dictionary}, and $f(\lambda)$ finally estimated from postprocessing this data.
Measurement schemes 2a and 2b of Fig.~\ref{fig:dictionary} are just the standard laboratory techniques of photon detection and Hanbury-Brown and Twiss interferometry.  Measurement scheme 2c represents an interferometer with variable path length that is used to estimate the derivative of the quantum field in the kinetic energy term $\langle \hat{T} \rangle$.  Length shifts on the mm scale correspond to ps values of $\epsilon_1$ and  $\epsilon_2$ in the estimation of  $\langle \hat{T} \rangle$; as these values are six orders of magnitude smaller than the relevant timescales of the experiment, they are considered sufficiently small.

 Obviously the test model chosen here is simple enough to admit a \emph{classical} simulation, which we carry out  for the experimental parameters of~\cite{Stute2012}.  Exploiting a simple gradient-descent method, we find the values of $\lambda = \{ g, \Omega, s \}$ that minimize $f(\lambda)$ for a given value of $v$. This procedure is repeated over a range of $v$ values of interest, and the corresponding optimized values of $\lambda$ are then used to compute quantities of interest, e.g., spatial-correlation functions as shown in Fig.~\ref{fig:pilot_study}. Remarkably, we find that just these three variational parameters $\lambda=(g,\Omega, s)$, when varied in the experimentally feasible parameter regime of \cite{Dubin2010, Stute2012} in the presence of losses, allow for a quantum simulation of Lieb-Liniger ground-state physics.  One expects that the ground-state approximation would improve by increasing the dimension of the auxiliary system and by allowing sufficiently general internal couplings and couplings to the field. In the context of atom-cavity systems, this can be done by making use of the rich level structure of atoms (i.e., Zeeman splittings) and making use of motional degrees of freedom.  For sufficiently complex intracavity dynamics a classical simulation will become unfeasible, and at the same time it becomes conceivable that such a simulator will outperform the best classical methods.



The reliability of a quantum simulator can be compromised by decoherence effects, as was recently emphasized in~\cite{Hauke2011}. Our simulation of the ion-cavity system includes both cavity decay (at rate $\kappa$) and decay of the ion due to spontaneous emission (at rate $\gamma$).
The cavity decay rate rescales the parameter $s$ linking measurement time and simulated space, and thus it can be considered as a
variational parameter itself.
We emphasize that cavity decay does not function as a decoherence channel in our scheme but is rather an essential element of the cMPS formalism. In contrast, spontaneous emission sets the limit for the timescale over which coherent dynamics can be observed.
For our present example, we can show that the regime of strong cooperativity $\mathcal{C} = g^2/\kappa\gamma \gtrsim 1$ is sufficient to allow a simulation of the Lieb-Liniger model despite detrimental losses. The Lieb-Liniger model exhibits nontrivial variations (Friedel oscillations) of the two-point correlation function on a length scale $\xi=\langle\hat\psi(x)\hat\psi^\dagger(x)\rangle^{-1}$. This length scale  in the simulated model translates to a time scale over which the stationary output field (as the simulating register) should exhibit similar nontrivial features. From the previously established scaling rule one finds that the required time scale is $\tau=\xi/s=\langle \hat E^{-}(t)\hat E^{+}(t)\rangle^{-1}$. By means of the cavity input-output relations, we relate the output photon flux to the mean intracavity photon number $\langle\hat E^{-}(t)\hat E^{+}(t)\rangle=\kappa\langle\hat a^\dagger \hat a\rangle$. In the bad cavity limit $\kappa \gg g$ the cavity mode can be adiabatically eliminated. In this case $\langle a^\dagger a\rangle\simeq (g/\kappa)^2$, such that $\tau=\kappa/g^2$. On the other hand, the characteristic decoherence time of the ion is determined by the spontaneous emission rate $2\gamma$. Beyond a time $1/2\gamma$ the second order correlation function $g^{(2)}$  will be trivial. We therefore  demand $\tau\lesssim 1/2\gamma$, which is equivalent to the requirement of a large cooperativity $\mathcal{C}\gtrsim 1$. For the exemplary case of the ion-cavity system considered above the cooperativity was indeed $\mathcal{C}\simeq 1.8$, see supplementary material. While equivalent conditions must be determined on a case-to-case basis, we expect that nontrivial quantum simulations in cavity QED will not be possible in the weak-coupling regime. Finally, there are overall losses associated with scattering and absorption in cavity mirrors, detection path optics, and photon-counter efficiency.  However, while these losses reduce the efficiency with which photon correlations are detected, they do not otherwise affect the system dynamics.

A natural question is when our scheme would provide a practical advantage over classical computers in the simulation of quantum fields. We expect this to be the case in particular for the simulation of fields with multiple components, or species of particles, with canonical field-annihilation operators $\hat{\psi}_\alpha(x)$, $\alpha = 1, 2, \ldots, N$. This situation arises in at least two settings: firstly, for vector bosons in gauge theories with gauge group $\textsl{SU}(N)$, and secondly, in the nonrelativistic setting of cold atomic gases  with multiple species. Variational calculations using cMPS fail in these settings, as the number of variational parameters must scale as $D\sim 2^N$. On the other hand, in a cavity-QED quantum simulation multiple output fields are naturally accessible via polarization or higher order cavity modes, and at the same time large effective Hilbert space dimensions can be achieved, e.g., with trapped ions or atoms. With $N \gtrsim 10$, substantial practical speedups are already expected with respect to the classical cMPS algorithm, which requires a number of operations scaling as $2^{3\times N}$.

The realisation that ground-state cMPS and the field states emerging from a cavity are connected can be exploited to characterise the correlations of the emitted light. Indeed, we obtain a simple criteria to determine when the correlations in the light field are nonclassical: if it turns out that the simulated hamiltonian is quadratic in the field operators; and (b) contains only ``ultralocal'' terms, i.e., no derivatives in the field operators, then the ground state is a trivial (i.e., gaussian) product state, and there would be no nonclassical correlations in the emitted light.

The output of a cavity-QED apparatus admits a natural interpretation as a variational class of quantum-field states. We have demonstrated that this allows an \emph{analogue} quantum simulation procedure for strongly correlated physics using current technology. This result opens up an entirely new perspective for all cavity-QED systems which exhibit sufficiently strong nonlinearities at the single-photon level. This includes not only optical cavities coupled to atoms, but also superconducting circuits with super-strong coupling to solid state quantum systems~\cite{Bozyigit2010}, as well as other nonlinear systems achieving a high optical depth without cavities, such as atomic ensembles exhibiting Rydberg blockade \cite{Pritchard}, or coupled to nanophotonic waveguides \cite{Vetsch2010, Goban2012}. Looking further afield, since the input-output and cMPS formalisms generalize in a natural way to fermionic settings~\cite{Sun1999,Search2002,Gardiner2004}, our simulation procedure might be applicable to cavity-like microelectronic settings involving fermionic degrees of freedom. We hope our work will inspire explorations of these promising directions.

\begin{acknowledgments}
We acknowledge helpful conversations and comments by Jens Eisert, Frank Verstraete, Ignacio Cirac, Bernhard Neukirchen, Jutho Haegeman, Konstantin Friebe, Jukka Kiukas and Reinhard Werner. This work was supported by the Centre for Quantum Engineering and Space-Time Research (QUEST), the ERC grant QFTCMPS, the Austrian Science Fund (FWF) through the SFB FoQuS (FWF Project No. F4003), the European Commission (AQUTE, COQUIT), the Engineering and Physical Sciences Research Council (EPSRC) and through the FET-Open grant MALICIA (265522).
\end{acknowledgments}

\begin{appendix}
  
\section{Supplementary Material}

\subsection{Continuous Matrix Product States and Cavity QED}

The matrix product state formalism has recently been generalized to the setting of quantum fields in \cite{Verstraete2010, Haegeman2010, Haegeman2010b} giving rise to \emph{continuous matrix product states} (cMPS). These states refer to one-dimensional bosonic fields with annihilation and creation operators $\hat\psi_\alpha(x)$ and $\hat\psi_\alpha^\dagger(x)$ (such that $[\hat\psi_\alpha(x),\hat\psi_\beta^\dagger(x)]=\delta_{\alpha\beta}\delta(x-y)$), and are defined as
\begin{align}\label{eq:Psi}
  |\Psi\rangle&=\mathrm{tr}_{aux}\{\hat U\}|\Omega\rangle,
\end{align}
where $|\Omega\rangle$ denotes the vacuum state of the quantum field and
\begin{align*}
  \hat U&=\mathcal{P}\exp\left(\int_{-\infty}^\infty dx \hat H_{\text{cMPS}}(x)\right).
\end{align*}
Here $\mathcal{P}$ denotes path ordering, and the (non Hermitian) Hamiltonian is
\begin{align}\label{eq:HcMPS}
\hat H_{\text{cMPS}}(x)=\hat Q\otimes\mathds{1}+\sum_\alpha \hat R_\alpha\otimes \hat\psi^\dagger_\alpha(x)
\end{align}
with $\hat Q$ and $\hat R_\alpha$ being $D\times D$ matrices acting on an auxiliary system of dimension $D$. $\mathrm{tr}_{aux}$ in Eq.~\eqref{eq:Psi} is the trace over this auxiliary system.

As was shown in \cite{Verstraete2010,Osborne2010} an equivalent representation of the state $|\Psi\rangle$ as defined in Eq.~\eqref{eq:Psi} is given by
\begin{align}\label{eq:Psi2}
  \hat\rho&\propto \lim_{x_0,x_1\rightarrow\infty}\mathrm{tr}_{aux}\left\{\hat U(x_0,x_1)|\Omega\rangle\langle\Omega|\otimes|\psi\rangle\langle\psi|\hat U^\dagger(x_0,x_1)\right\}
\end{align}
where
\begin{align*}
  \hat U(x_0,x_1)&=\mathcal{P}\exp\left(\int_{x_0}^{x_1} dx \hat H_{\text{cMPS}}(x)\right),
\end{align*}
and $|\psi\rangle$ is an arbitrary state of the auxiliary system, which plays the role of a boundary condition at $x_0$. The two states in \eqref{eq:Psi} and \eqref{eq:Psi2} are equivalent in the sense that they give rise to identical expectation values for normal- and position-ordered expressions of field operators $\langle\hat\psi^\dagger(y_1)\ldots\hat\psi^\dagger(y_n)\hat\psi(y_{n+1})\ldots \hat\psi(y_m)\rangle$, see \cite{Verstraete2010,Osborne2010}. Note that the trace in \eqref{eq:Psi2} is a proper partial trace over the auxiliary system (in contrast to the trace in \eqref{eq:Psi}).

Consider, on the other hand, a cavity with several relevant modes described by annihilation/creation operators $\hat a_\alpha,\,\hat a^\dagger_\beta$ (such that $[\hat a_\alpha,\hat a^\dagger_\beta]=\delta_{\alpha\beta}$) which are coupled to some intracavity medium via a system Hamiltonian $\hat H_{sys}$. Moreover, each cavity is coupled to a continuum of field modes ($[\hat a_\alpha(\omega),\hat a^\dagger_\beta(\bar\omega)]=\delta(\omega-\bar\omega)\delta_{\alpha\beta}$) through one of its end mirrors. (The generalization to the case of double-sided cavities, or ring cavities with several outputs per mode is immediate.) The total Hamiltonian for the cavities, the intracavity medium, and the outside field is
\begin{align*}
\hat H_{cQED}(t)&=\hat H_{sys}\otimes\mathds{1}\\
&\quad+i\sum_\alpha \int d\omega\sqrt{\frac{\kappa_\alpha(\omega)}{2\pi}}\left(\hat a_\alpha \otimes \hat a_\alpha^\dagger(\omega)e^{-i\omega t}-\mathrm{h.c.}\right).
\end{align*}
This Hamiltonian is written in an interaction picture with respect to the free energy of the continuous fields, and it is taken in a frame rotating at the resonance frequencies of the cavities. In the interaction picture and rotating frame each integral extends over a band width of frequencies around the respective cavity frequencies. In the optical domain the Born-Markov approximation, which assumes $\kappa_\alpha(\omega)=\mathrm{const.}$ in the relevant band width, holds to an excellent degree. In this case it is common to define time-dependent operators $\hat E^+_\alpha(t)=\int d\omega/\sqrt{2\pi}\,\hat a_\alpha(\omega)\exp(-i\omega t)$, which fulfill $[\hat E^+_\alpha(t),\hat E^-_\beta(t')]=\delta(t-t')$. As in the main text, these operators correspond to the electric field components at the cavity output, and are defined such that $\langle\hat E^{-}(t)\hat E^{+}(t) \rangle$ denotes the flux of photons per second. The Hamiltonian then is
\begin{align*}
\hat H_{cQED}(t)&=\hat H_{sys}\otimes\mathds{1}+i \sum_\alpha \sqrt{\kappa_\alpha}\left(\hat a_\alpha \otimes \hat E_\alpha^-(t)-\mathrm{h.c.}\right).
\end{align*}
If the outside field modes are in the vacuum, standard quantum optical calculations \cite{Gardiner2004a} show that this Hamiltonian is equivalent to an \emph{effective} non-Hermitian Hamiltonian
\begin{align*}
\hat H_{cQED}(t)&=\hat H_{sys}\otimes\mathds{1}-i\sum_\alpha\frac{\kappa_\alpha}{2}\hat a^\dagger_\alpha \hat a_\alpha\otimes\mathds{1}+i \sum_\alpha \sqrt{\kappa_\alpha} \hat a_\alpha \otimes \hat E_\alpha^-(t).
\end{align*}
When this is compared to Eq.~\eqref{eq:HcMPS} the identification of the formalism of cMPS and cavity QED is immediate. If we assume that the field and the cavity system is in a state $|\Omega\rangle\otimes|\psi\rangle$ at some initial time $t_0$ the final state of the field outside the cavity at time $t_1$ is given by expression \eqref{eq:Psi2}, when we identify $x=s\,t$, and $\hat H_{\text{cMPS}}(x)=-i\hat H_{cQED}(x/s)$, that is
\begin{align}\label{eq:identification}
  \hat \psi(x)&= \frac{1}{\sqrt{s}}\hat E^+_\alpha(t), &
  \hat R_\alpha&=\sqrt{\frac{\kappa_\alpha}{s}}\hat a_\alpha, &
  \hat Q&= -\frac{i}{s}\hat H_{sys}-\sum_\alpha\frac{\hat R^\dagger_\alpha \hat R_\alpha}{2},
\end{align}
with an arbitrary scaling factor $s$. Therefore, the state of the output modes of a cavity is always a continuous matrix product state. Formally these states are cMPS with an infinite-dimensional auxiliary system $D\rightarrow\infty$. However, due to energy constraints, the dimensions of the cavity system are effectively finite. The relevant dimension of the cavity Hilbert space then sets the dimension $D$ of the auxiliary system in the cMPS formalism.


\subsection{Quantum Variational Algorithm}

As an exemplary test case we demonstrated that the cavity QED system comprising a single trapped atom strongly coupled to a single high-finesse cavity mode  is capable of simulating the ground-state physics of the Lieb-Liniger model. The atom-cavity system is described (in a suitable rotating frame) by the on-resonance Jaynes-Cummings Hamiltonian
\begin{align}
H_{sys} = g(\hat \sigma^{+}\hat a + \hat \sigma^{-}\hat a^{\dagger}) + \Omega(\hat\sigma^{+} + \hat\sigma^{-}),
\label{HJC1}
\end{align}
where $\hat\sigma^{+}$ is the atomic raising operator and $\hat a$ is the cavity-photon annihilation operator, $g$ the atom cavity coupling,  and $\Omega$ the laser drive.  Photons leak out of the cavity with leakage rate $\kappa$, and it is assumed that, in a real experiment, this output light can be measured by various detection setups.

The Lieb-Liniger model, with Hamiltonian
\begin{align*}
{\cal \hat H}  &= \int_{-\infty}^\infty (\hat{T}+\hat{W}+\hat{N})dx \\
&= \int_{-\infty}^\infty \Big[ \frac{d \hat \psi^\dagger(x)}{dx} \frac{d \hat \psi(x)}{dx}  + v~ \hat \psi^\dagger(x)\hat \psi^\dagger(x) \hat\psi(x) \hat\psi(x)\\
&\hspace{5cm} - \mu \hat \psi^\dagger(x) \hat \psi(x) \Big]
dx,
\end{align*}
describes hard-core bosons with contact interaction of strength $v$.  We performed variational optimizations for a range of values for $v$. We did this using a simple gradient-descent minimization of the average energy density  $f(\lambda) = \langle \Psi(\lambda)|\hat{T}+\hat{W}+\hat{N}|\Psi(\lambda)\rangle$. Using the parameters that minimize  $f( \lambda)$  we then calculate other quantities of interest, such as correlation functions for the simulated ground state field. We have focussed on a gradient-descent algorithm for clarity; in practice a more sophisticated optimization procedure using the time-dependent variational principle, or conjugate gradients, could be used.

Our variational parameters $ \lambda = \left( g, \Omega, s \right)$ enter $f(\lambda)$ as follows. The expectation value of the energy density in the Lieb-Liniger model is
\begin{align*}
f(\lambda; v,\mu) = \langle \hat{T} \rangle +\langle \hat{W} \rangle + \langle \hat{N} \rangle
\end{align*}
where each term may be written in terms of the experimentally observed correlation functions via the correspondence $\hat\psi(x) = \hat E^+(t)/\sqrt{s}$, giving
\begin{align*}
\langle \hat{T} \rangle &= \lim_{\epsilon_1,\epsilon_2\rightarrow 0} \frac{1}{s^3\epsilon_1\epsilon_2}\left(g^{(1)}(t+\epsilon_1,t+\epsilon_2)-g^{(1)}(t+\epsilon_1,t)\right.\\
&\left.\quad-g^{(1)}(t,t+\epsilon_2)+g^{(1)}(t,t)\right) \,, \\
\langle \hat{W}\rangle & = \frac{v}{s^2} g^{(2)}(t,t) \,, \\
\langle \hat{N} \rangle & = -\frac{\mu}{s} g^{(1)}(t,t) \,,
\end{align*}
where $g^{(1)}(t_1,t_2)=\langle \hat E^{-}(t_1)\hat E^{+}(t_2)\rangle$ and $g^{(2)}(t_1,t_2)=\langle \hat E^{-}(t_1)\hat E^{-}(t_2)\hat E^{+}(t_2)\hat E^{+}(t_1)\rangle$. In an experimental simulation, these correlation functions would be measured directly in the laboratory, and the results would be fed back into a classical computer performing the optimization algorithm. However, for the purposes of our proof-of-principle simulation, we calculate the correlation functions directly by means of the input-output formalism. Making use of the cavity input-output relation $\hat E^{+}(t) = \hat E^+_{\mathrm{(in)}}(t) + \sqrt{\kappa}\hat a(t)$, where $\hat E^+_{\mathrm{(in)}}(t)$ denotes the field impinging on the cavity at time $t$ (which is assumed to be in the vacuum state), the expectation value may be written \cite{Verstraete2010,Osborne2010}

\begin{align*}
 f(\lambda; v,\mu)  =   \mathrm{tr} \left\{ \left(\left[\hat Q,\hat R\right]\right)^\dagger \left[\hat Q,\hat R\right] \hat \rho_{\tiny \mbox{ss}} \right\}  &+  v~ \mathrm{tr} \left\{ \left(\hat R^\dagger\right)^2\hat R^2 ~\hat\rho_{\tiny \mbox{ss}}\right\}\\
  &- \mu~\mathrm{tr} \left\{ \hat R^\dagger \hat R \hat\rho_{\tiny \mbox{ss}}\right\}
 \end{align*}
where $\hat R$ and $\hat Q$ are defined in Eq.~\eqref{eq:identification}, and $\hat\rho_{\tiny \mbox{ss}}$ is the unique steady state of the atom-cavity system, satisfying
\begin{align*}
\frac{d \hat\rho_{\tiny \mbox{ss}} }{dt} = -i \left[  \hat H_{\tiny\mbox{sys}}, \hat\rho_{\tiny \mbox{ss}}\right] + \kappa \hat a \hat\rho_{\tiny \mbox{ss}} \hat a^\dagger -\frac{\kappa}{2} \left[ \hat a^\dagger \hat a, \hat\rho_{\tiny \mbox{ss}}\right]=0.
\end{align*}
Note that above, we write $f(\lambda; v,\mu)$ to highlight the dependence of $f$ on $v$ and $\mu$. Hereafter we set $\mu=1$ and minimize $f(\lambda; v,\mu)$ for a range of different values of $v$. Solutions for other values of $\mu$ can be obtained by means of a scaling transformation, as described in \cite{Verstraete2010,Osborne2010}.

The general outline of the algorithm to determine our optimum values of the variational parameters $\lambda$, for a given choice of $v$ and $\mu$, is thus:
\vspace{1em}
\begin{algorithmic}                    
\STATE initialize $\textit{tol}$ (tolerance), and $\epsilon$ (step size)
\STATE initialize $\lambda$ to an arbitrary value \\
\REPEAT
\STATE set $\lambda^{\prime} = \lambda$\\
\STATE calculate $\nabla f(\lambda; v,\mu)$\\
\STATE update experimental parameters as $\lambda \leftarrow \lambda - \epsilon\nabla f(\lambda; v,\mu)$\\
\STATE calculate $f(\lambda^{\prime}; v,\mu)$ and $f(\lambda; v,\mu)$\\
\UNTIL{$|f(\lambda; v,\mu) -  f(\lambda^{\prime}; v,\mu)| < tol$}
\end{algorithmic}
\vspace{1em}
Note that in our simulation, $\nabla f(\lambda; v,\mu)$ is estimated numerically by evaluating $f(\lambda+\Delta_i; v,\mu)$ for small values of $\Delta_i$, while in an experiment, $\nabla f(\lambda; v,\mu)$ is found with the aid of measurements of $\langle \hat{T} \rangle$, $\langle \hat{W}\rangle$, and $\langle \hat{N} \rangle$.

\end{appendix}

\end{document}